# Digital-Analog Transmission Framework for Task-Oriented Semantic Communications

Yuzhou Fu, *Student Member, IEEE*, Wenchi Cheng, *Senior Member*, *IEEE*, Wei Zhang, *Fellow*, *IEEE*, and Jingqing Wang, *Member, IEEE*

*Abstract*—Task-Oriented Semantic Communication (TOSC) has been considered as a new communication paradigm to serve various samrt devices that depend on Artificial Intelligence (AI) tasks in future wireless networks. The existing TOSC frameworks rely on the Neural Network (NN) model to extract the semantic feature from the source data. The semantic feature, constituted by analog vectors of a lower dimensionality relative to the original source data, reserves the meaning of the source data. By conveying the semantic feature, TOSCs can significantly reduce the amount of data transmission while ensuring the correct execution of the AI-driven downstream task. However, standardized wireless networks depend on digital signal processing for data transmission, yet they necessitate the conveyance of semantic features that are inherently analog. Although existing TOSC frameworks developed the Deep Learning (DL) based *analog approach* or conventional *digital approach* to transmit the semantic feature, but there are still many challenging problems to urgently be solved in actual deployment. In this article, we first propose several challenging issues associated with the development of the TOSC framework in the standardized wireless network. Then, we develop a Digital-Analog transmission framework based TOSC (DA-TOSC) to resolve these challenging issues. Future research directions are discussed to further improve the DA-TOSC.

## I. INTRODUCTION

With the advent of various emerging applications, such as smart monitoring, autonomous driving, and smart furniture, massive smart devices rely on the Artificial Intelligence (AI) driven downstream tasks, such as detection and prediction, to serve users within a tolerated delay. The massive connection of the smart device further raises high demand for low-latency and high-reliable wireless transmission [1]. Classic semantic communication, initially introduced by Weaver [2], as an extension of Shannon information theory, aims to guide the receiver towards intended goal by conveying the meaning of data rather than accurate data transmission, thus improving further communication efficiency. Due to the limitations of computers and hardware, the *semantic* aspects of communication are regarded as *irrelevant* to the engineering problem, leading to its neglect in the discourse. The research on semantic communication has been ignored in previous decades. The emergence of task-oriented communication paradigms necessitates a critical reassessment of the traditional frameworks predicated on Shannon's theory. This calls for a renewed focus on semantic communication frameworks [3], prompted by the potential enhancements they offer to communication efficacy. Recently, International Telecommunication Union (ITU) published the latest vision on the semantic-aware networking, identifying semantic communication as a promising candidate to support various emerging applications in future networks [4].

Benefiting from the development of Deep Learning (DL), many DL based Task-Oriented Semantic Communication (TOSC) frameworks have been proposed for serving downstream tasks [5]–[9]. Depending on the Neural Network (NN) models, these TOSC frameworks analyze the intention of the downstream task and then extract the semantic feature from the source data according to the analytical result. The semantic feature, constituted by analog vectors of a lower dimensionality relative to the original source data, reserves the meaning of the source data. Also, the correct execution of the downstream task depends on original meaning of the data rather than high-quality data reconstruction. By conveying the semantic feature instead of the source data, TOSC can significantly reduce the amount of data transmission while ensuring the correct execution of the downstream task. For example, the authors of [5] proposed a joint perception and decision scheme, which learns optimal semantic compression scheme for classification task and aims to balance transmission latency and classification accuracy under different channel conditions. In [6], the authors proposed Reinforcement Learning (RL) based semantic bit allocation model, which is used to guide the semantic encoding with adaptive quantization, thus achieving adaptive semantic compression for TOSC. Apart from compressing the semantic feature according to the task intentions, the authors of [7], [8] proposed efficient semantic extraction based TOSCs, which only extract task-related semantic features for transmission instead of full semantic features, thus further improving the transmission efficiency of task-oriented communication. The task-related semantic feature refers to a subset of the semantic feature that significantly contributes to the correct execution of downstream tasks [8]. The contribution of each semantic feature can be naturally represented by model weights of the downstream task. In practice, the transmitter can obtain pre-trained model weights to extract the task-related semantic feature. Although the feasibility of TOSC framework has been validated, the deployment of TOSC frameworks in standardized wireless networks presents many new challenges.

One of the significant challenges lies in deploying the TOSC

Yuzhou Fu, Wenchi Cheng, and Jingqing Wang are with State Key Laboratory of Integrated Services Networks, Xidian University, Xi'an, 710071, China (e-mails: fyzhouxd@stu.xidian.edu.cn; wccheng@xidian.edu.cn; jqwangxd@xidian.edu.cn).

Wei Zhang is with School of Electrical Engineering and Telecommunications, the University of New South Wales, Sydney, Australia (e-mail: w.zhang@unsw.edu.au).

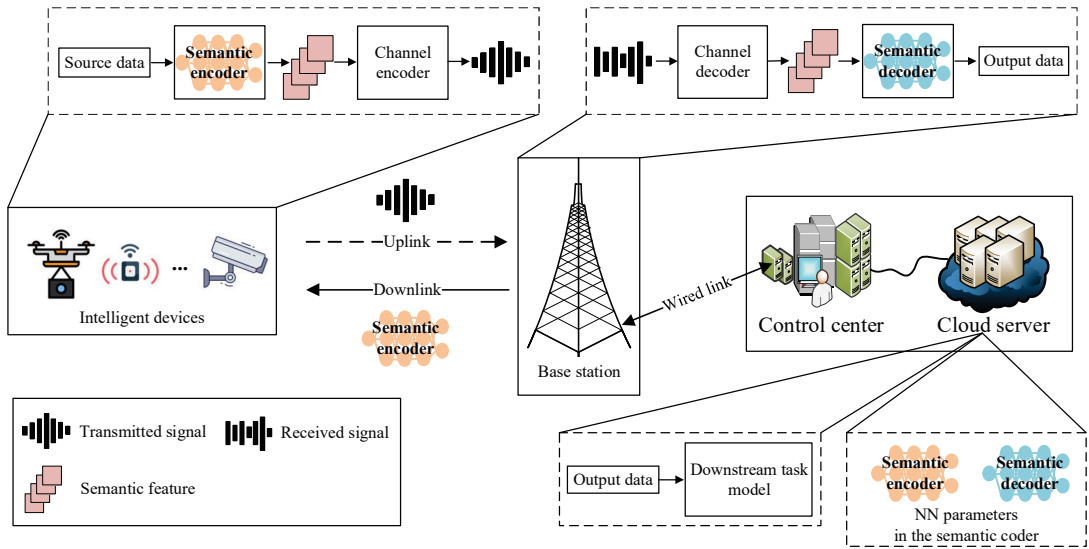

Fig. 1: The architecture of TOSC-enabled wireless network.

framework in standardized wireless networks. These networks depend on digital signal processing for data transmission, yet they necessitate the conveyance of semantic features that are inherently analog. According to the different signal processing approaches to convey the semantic feature, we can divide the existing TOSC framework into *analog approach* [5], [6], [10] and *digital approach* [8], [9]. The *analog approach* adopts DL based Joint Source-Channel Coding (JSCC) to learn noise resilient channel input, which directly maps analog semantic feature to the channel input for wireless transmission. It has significant advantages in improving the robustness in poor channel conditions but fails to fully use the Digital Signal Processing (DSP) system of mature wireless networks. In addition, the analog approach needs to map the semantic feature vector with full-resolution constellation into a finite space, which makes it possible to implement in the existing radio frequency system [10] but leads to the *saturation effect*. The *saturation effect* indicates that the performance of TOSC cannot achieve the upper limit of its design as the channel SNR increases. On the other hand, the *digital approach* is to quantize analog semantic feature as digital signal in order to adopt conventional digital signal processing, such as source compression, channel coding, and carrier modulation. Since *digital approach* is based on classic separated source and channel coding scheme, the virtue of the robust transmission is lost. Moreover, *digital approach* inevitably experiences more severe *cliff effect* and *saturation effect* than the *analog approach* due to the loss of information caused by quantification. The *cliff effect* means that the performance of semantic communication degrades drastically when the channel Signal-to-Noise Ratio (SNR) falls beneath the designed SNR. In addition to the above outstanding problems, the standardized wireless networks will also cause many problems when introducing TOSC framework in the aspects of the semantic encoder updating and resource utilization. Inspired by Digital-Analog (DA) transmission framework [11], [12], we propose the DA transmission framework for TOSC (DA-TOSC), which integrates the DL based JSCC scheme and widely adopted digital systems prevalent in existing networks to maximize the performance of TOSC, thus achieving effective deployment of TOSCs.

In this article, we fist present an overview of the general architecture of TOSC-enabled wireless network and study several challenging problems in deploying TOSC in standardized wireless networks. Then, we develop DA-TOSC to resolve these challenging problems. In order to illustrate validity, an exemplary use case of proposed DA-TOSC is presented. Future research directions for enhancing system performance of TOSC-enabled wireless networks with Digital-Analog (DA) transmission are also discussed.

## II. Challenging Problems in Effectively Deploying TOSC into Standardized Wireless Networks

Figure 1 shows the architecture of TOSC-enabled wireless network, where the intelligent devices adopt uplink to convey semantic feature for the base station. Under the TOSC framework, both semantic encoder and the semantic decoder consist of the neural networks, where the semantic encoder is responsible for extracting semantic feature and the semantic decoder is responsible for estimating an output data to initiate the downstream task. Also, the channel encoder and the channel decoder can adopt digital coding scheme or the NN based analog coding scheme depending on the transmission approach. The base station considered as the wireless access point is equipped with the semantic decoder and the channel decoder to estimate an output data according to the received signal. In practice, the emerging networks, such as Internet of Things (IoT) and smart factory, will deploy the cloud server-enabled control center to perform various downstream tasks in the background. The cloud service is still responsible for re-training semantic encoder and semantic decoder for the intelligent device and the base station in order to match current data type [13]. The control center can connect to the base station through the wired link to efficiently send the output data for the downstream task. However, the existing *analog approach* and *digital approach* do not facilitate the optimal utilization of the digital system nor do they maximize the performance capabilities of TOSC, and thus they are not the effective deployment for the TOSC. As illustrated in Fig. 2, the challenging problems of the existing deployment scheme (namely, analog approach and digital approach) contain overlooking widely adopted digital systems, poor performance of TOSC, and limited performance of TOSC. Then, we propose digital-analog transmission framework for the TOSC to achieve the effective deployment and it also lead to three challenging researches.

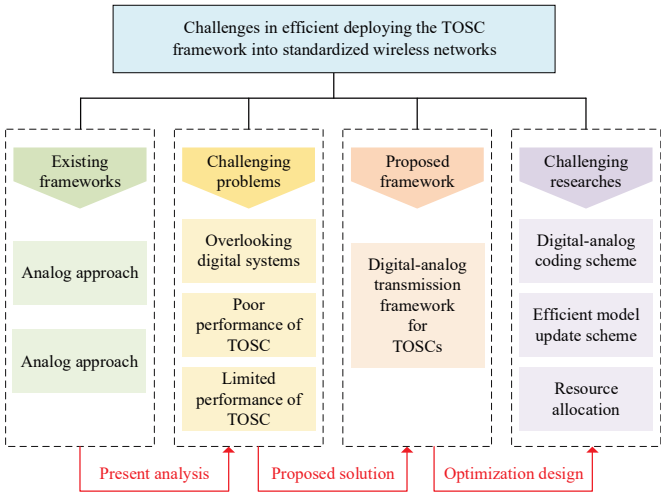

Fig. 2: The challenges in effectively deploying TOSC into standardized wireless networks.

### A. Overlooking Widely Adopted Digital Systems

In the existing TOSC framework, the *analog approach* relies on advanced NNs model to enable the end-to-end communication system, which can effectively improve system robustness in poor channel conditions but overlooks the integration of widely adopted digital systems prevalent in existing networks, failing to consider these systems comprehensively in the design process. In fact, the mature DSP schemes can be used to efficiently deliver side information and update model parameters, thus maximizing the performance of TOSCs. Moreover, the digital system will also be an important part of future wireless networks from an engineering point of view. Thus, the effective use of digital system is a challenging problem in deploying TOSC in standardized wireless networks.

### B. Poor Performance of TOSC

On the other hand, the *digital approach* can quantize the analog semantic feature vector as digital signal, which can be transmitted with mature DSP technologies but leads to the loss of semantic information. Also, the *digital approach* is based on classic separated source and channel coding scheme, which exhibits limited poor robustness to the channel noise, especially in low SNR conditions. Thus, the poor noise robustness of the *digital approach* can further cause more semantic information loss. Due to the loss of semantic information, the *digital approach* inevitably experiences severe *cliff effect* and *saturation effect*. The two negative effects can severely degrade the performance of TOSC.

### C. Limited Performance of TOSC

In fact, both *analog approach* and *digital approach* of the TOSC frameworks rely on the NNs model to extract the semantic feature from the source data and to predict the information required for downstream task according to the received signal. Thus, the performance of the TOSC framework is significantly influenced by the NNs model integrated into it. In practice, there may be some factors that prevent the model from achieving the expected performance, thereby limiting the performance of TOSC. For example, the NNs model needs to map the full-resolution semantic feature extracted from the source data into finite constellation for easy wireless transmission using existing radio frequency system. The trained NNs model is compressed into a lightweight model, which is for easy deployment on mobile devices but inevitably causes performance degradation. The NNs model is not updated on time, resulting in semantic analysis errors. Thus, the limited performance of TOSC caused by the NNs model is a common challenging problem for *analog approach* and *digital approach*.

Inspired by DA transmission framework, we develop the DA transmission framework for TOSC (DA-TOSC), which integrates the DL based JSCC scheme and widely adopted digital systems prevalent in existing networks to maximize the performance of TOSC. However, the DA-TOSC presents many challenging researches, such as the digital-analog coding scheme, efficient model update scheme, and the resource allocation.

### A. Digital-Analog Coding Scheme

The DA-TOSC framework needs to adopt digital-analog coding scheme, which combines the DL based analog coding and the digital coding in order to maximize the semantic fidelity. Also, the code rate optimization plays a crucial role in enhancing the performance of digital-analog coding scheme [11]. It is a tradeoff problem in terms of the number of digital-analog transmitted symbols, channel conditions, and the performance of downstream task. For example, when the SNR is better, we can use the digital coding branch to transmit more symbols for conveying side information, which focuses on improving the data fidelity rather than semantic fidelity to approach the performance upper limit of downstream task. If the SNR is poor, the analog coding branch is required to provide transmission robustness, thus ensuring the performance lower bound of the TOSC for the correct execution of the downstream task.

### B. Efficient Model Update Scheme

The existing wireless networks depend on digital control signals to update the NNs model. However, the semantic encoder contains a large number of high-precision NN parameters, which will create massive downlink control signals to occupy the channel resources for uplink semantic feature transmission. The reliance on digital signals for model updating may result in a scenario where the receiver is unable to acquire an adequate amount of semantic features necessary for the downstream task within an acceptable task delay tolerance. This insufficiency could lead to the failure of the downstream task. Based on the digital-analog transmission framework, we need to develop efficient model update scheme, which can fully use digital transmission and analog transmission to significantly reduce downlink overhead.

### C. Resource Allocation

In DA-TOSC framework, resource allocation plays a pivotal role in achieving the desired performance of TOSC-enabled wireless networks. The allocation of power and coding rate should be jointly considered in resource allocation, since they not only determine the amount of information for digital transmission and that for analog transmission but also affects the semantic fidelity of the output data. Several research studies have explored joint resource allocation schemes for the DA transmission framework to strike performance balance between digital transmission and analog transmission [14]. However, these resource allocation schemes focus on maximizing transmission fidelity rather than semantic fidelity, and thus are not valid for the TOSC. Thus, how to develop efficient resource allocation scheme for the DA-TOSC framework is a challenging research.

## III. DIGITAL-ANALOG TRANSMISSION FRAMEWORK FOR TASK-ORIENTED SEMANTIC COMMUNICATIONS

Figure 3 shows the framework of the proposed DA-TOSC for wireless networks. In practice, all data processing in the equipment are conducted by digital signal processing, where the continuous-amplitude (analog) signals consist of floating-point variables with precision depending on the digital processor. Then, only channel input signals are true analog mappings. As shown in Fig. 3, the data processing of uplink semantic transmission and downlink parameters transmission consists of digital and analog branches. The analog branch is responsible for processing analog signals, such as semantic features and floating-point parameters, thus achieving robust and full-precision transmission. Also, the digital branch is responsible for processing discrete signals and integer parameters with Distributed Source Coding (DSC), which can enhance semantic fidelity in uplink semantic transmission and reduce control overhead in downlink parameters transmission, respectively. Based on the multiplexer, the analog branch output and digital branch output can be mapped to same wireless channel with the orthogonally multiplexing, such as frequency division or antenna diversity. Then, the receiver can easily distinguish the received signals of the digital branch and the analog branch according to predetermined radio frequency or receiving antenna. The multiplexer is widely integrated into existing wireless networks to achieve orthogonal transmission. Thus, the DA-TOSC framework introduces a minimal increase in system complexity when compared to the digital approach and analog approach frameworks. This incremental complexity primarily originates from the integration of a multiplexer within the radio unit. Based on the performance gain of DA transmission framework, the DA-TOSC can adopt lightweight neural network model without losing system performance and executes the digital branch and the analog branch in parallel, thus reducing the system costs. In the following, we present several schemes for the DA-TOSC framework.

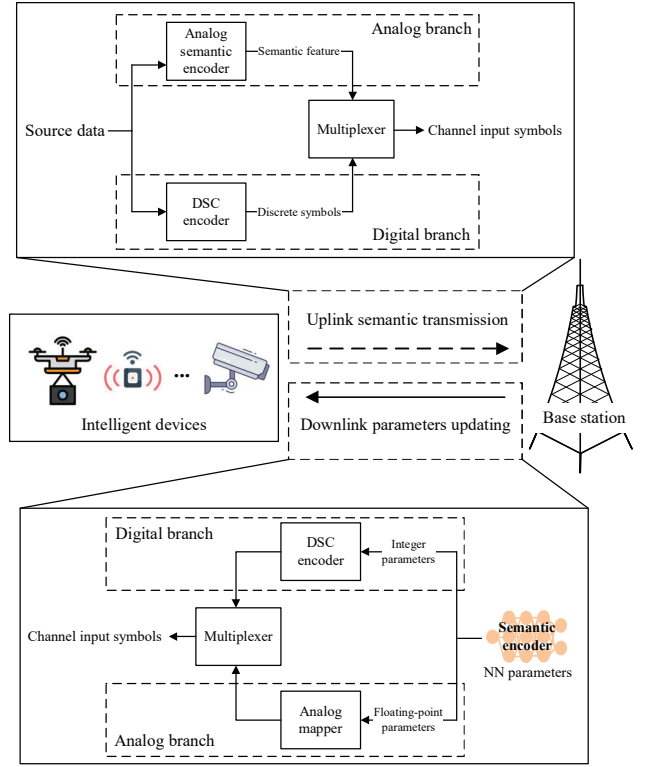

Fig. 3: The framework of the proposed DA-TOSC for wireless networks.

### A. DA Transmission Framework Based JSCC Scheme

Figure 4 shows an architecture of the proposed DA-JSCC. At the transmitter, the DL based semantic encoder is responsible for extracting task-related semantic feature from the source data and directly maps the semantic feature to the channel input signal. The DSC encoder adopts conventional channel coding to encode the source data and only outputs the parity bits as the side information to further improve the semantic fidelity. Then, the parity bits is modulated to channel input signal. The channel input signals of the semantic encoder and the DSC encoder are fed to the multiplexer, thus achieving multiplex transmission with the multiplexing technology. The code rate controller is responsible for determining analog code rate and digital code rate in terms of task requirements and total rate constraint. With DA transmission framework, the code rate is defined as the ratio between the number of source symbols and the length of channel input signals. In particular, the code rate controller is designed to first allocate the analog code rate required for semantic encoder in order to ensure the performance lower bound of the TOSC. Then, the digital encoding with multiple code rates is adopted to further enhance the semantic fidelity under total code rate constraint. At the receiver, the received signals are demultiplexed and conveyed to the semantic decoder and the DSC encoder. Specifically, the semantic decoder can transform received analog signals into some Intermediate Estimation Outputs (IEOs) and an Ultimate Estimation Output (UEO), where IEOs and UEO are obtained from the intermediate NN layer and the last NN layer, respectively. Then, the IEOs and UEO are respectively used as soft information and hard information for channel decoding in the DSC decoder, thus enhancing the semantic fidelity of the output data.

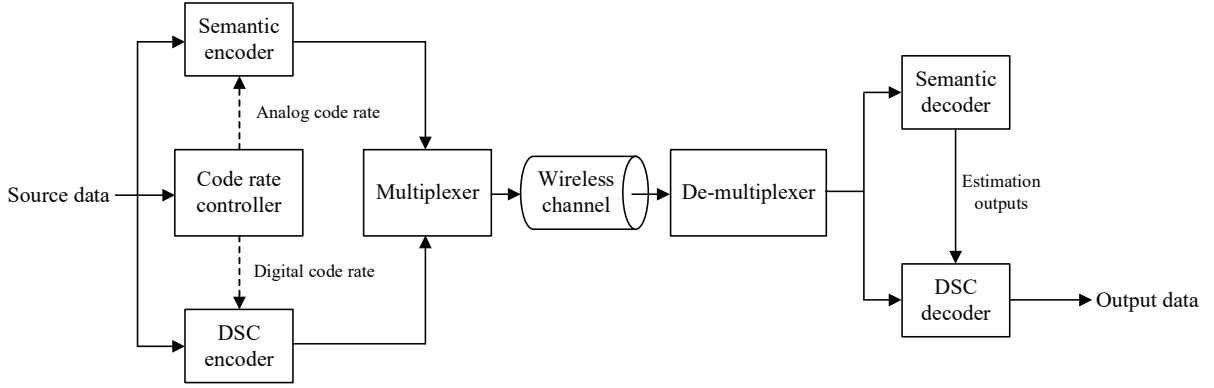

Fig. 4: The architecture of DA transmission framework based JSCC scheme.

### B. DA Transmission Framework Based Semantic Encoder Updating Scheme

To solve the massive downlink control overhead of updating Semantic Encoder (SE) in the DA-TOSC-WN, the DA transmission framework based Semantic Encoder Updating (DA-SEU) scheme can be used for downlink parameters transmission, as shown in Fig. 3. In order to achieve lightweight deployment, the parameters in the semantic encoder are comprised of floating-point parameters and integer parameters with a precision depending on the importance of NN layers. With DA-SEU scheme, the floating-point parameters are entered to the analog mapper for subsequent analog transmission, thus updating high-precision parameters. In digital branch, the proposed DA-SEU scheme still adopts DSC, which can fully utilize the correlation between the updated parameters and outdated parameters to convey integer parameters while reducing downlink control overhead. The DSC encoder is to encode the integer parameters of updated semantic encoder and only output its parity bit. The base station only transmits a small number of parity bitstream to the device with digital transmission. The device can adopt channel decoding to correct the outdated integer parameters, thus updating integer parameters in the semantic encoder. With the proposed DA-SEU scheme, the massive downlink control overhead problem caused by updating SE can be solved in DA-TOSC-WN.

### C. DA Transmission Framework Based Efficient Resource Allocation Scheme

Based on the DA-TOSC framework, we attempt to perform joint coding rate and power allocation scheme to optimize system distortion, which is the weighted sum of the analog distortion and the digital distortion. The analog distortion, which is defined as the Mean Square Error (MSE) between the semantic feature of the source data and the semantic feature of output data. The digital distortion is defined as the MSE between the source data and the output data. Thus, the system distortion is influenced by the SNR and the amount of information that is transmitted through the analog branch and the digital branch. Moreover, we set two weighted factors, such as $\lambda \in (0,1)$ and $(1-\lambda)$, for the analog distortion and the digital distortion, respectively. With setting the high weighted factor to the analog distortion, the resource allocation scheme tends to pay more resources to the analog branch in order to achieve efficient semantic transmission, and the digital branch focuses on further improving the semantic fidelity with limited power and coding resources. Hence, we formulate the joint allocation problem as an expected system distortion minimization problem, thus maximizing the semantic fidelity of the DA-TOSC framework. Then, we can adopt the iterative gradient algorithm or the greedy algorithm to find the optimal solution.

## IV. EXEMPLARY USE CASE OF PROPOSED DA-TOSC FRAMEWORK

In order to verify the effectiveness of the proposed DA-TOSC framework, the semantic encoder adopts the Scalable Extraction based Semantic Coding (SE-SC) scheme [7], which is an efficient semantic extraction for TOSCs. Then, the *analog approach* and *digital approach*, serving as comparative schemes, both adopt the SE-SC scheme to extract the semantic feature but transmit the semantic feature depending on the DL based analog coding scheme and conventional digital coding scheme, respectively.

Fig. 5 shows the semantic fidelity of proposed DA-TOSC, *analog approach*, and *digital approach* versus average SNR, where we adopt quasi-static Rayleigh fading channel, object detection as downstream task, and ImageNet-1K as the dataset in this article. Thus, the semantic fidelity can be defined as the mean Average Precision (mAP) to measure the performance of detection task. It can be observed from Fig. 5 that the semantic fidelities of *analog approach* and *digital approach* schemes remain nearly constant and present *saturation effect* when average SNR is better than 10dB. This is because *analog approach* and *digital approach* schemes have reached their design limits of keeping data information. Due to lossy semantic compression, some detailed information of the source data fail to be restored, making them impossible to further improve semantic fidelity even under error free transmission conditions. However, the proposed DA-TOSC is designed to spend a small portion of code for digital DSC to restore lost information, thus achieving the *graceful improvement* of the semantic fidelity. It also can be observed from Fig. 5 that the semantic fidelity of *digital approach* scheme drastically degrades and presents *cliff effect* when average SNR drops beneath 10dB, due to the shortage of digital transmission. However, the proposed DA-TOSC and *analog approach* schemes have shown excellent noise-robustness.

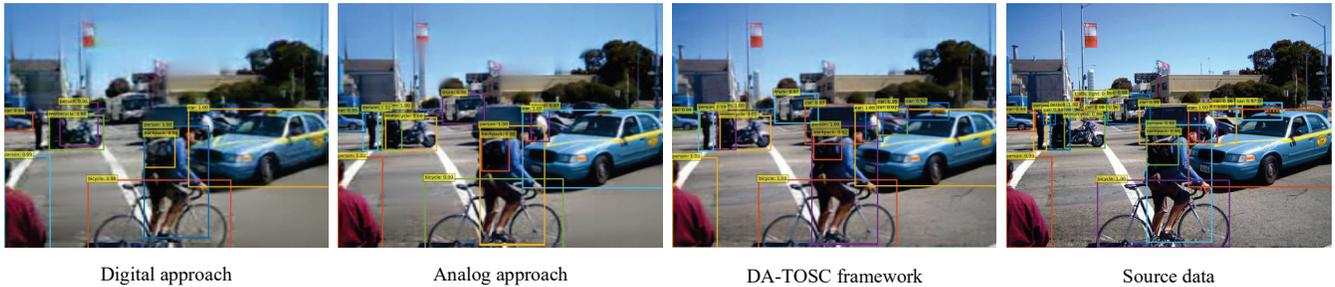

Fig. 6: A visual example of proposed DA-TOSC, digital approach, analog approach, and source data in the object detection task.

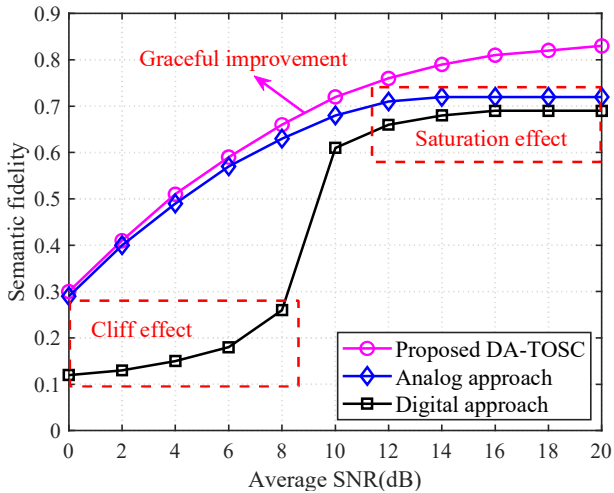

Fig. 5: Semantic fidelity of proposed DA-TOSC, digital approach, and analog approach versus average SNR.

Figure 6 shows a visual example of proposed DA-TOSC framework, digital approach, analog approach, and source data in the object detection task, where the SNR is set to 8dB. It can be observed from Fig. 6 that the *analog approach* scheme can only identify close and large objects, but fails to identify many distant objects even with good pixel-wise reconstruction. Since the *digital approach* scheme experiences serve *cliff effect*, the received semantic features are severely distorted and cannot efficiently support the object detection. It also can be observed from Fig. 6 that the *analog approach* scheme can identify most objects except for overlapping vehicles in the distance. This is because that *analog approach* has reached the limit of its semantic fidelity. Based on the DA transmission framework, our proposed DA-TOSC framework can further improve the semantic fidelity. As shown in Fig. 6, the DA-TOSC framework can identify almost all objects compared to the source data, thus more effectively serving the downstream task than the *analog approach* and *digital approach* schemes.

## V. Future Research Directions

Under the DA transmission framework, academic and industrial researchers should focus on a number of future research directions for DA-TOSC. The wide range of research directions contain physical (PHY) layer, application layer, and heterogeneous network. In the following, a brief discussion for future research directions in the DA-TOSC is given.

The energy efficiency also is the main objective in the PHY layer. With DA transmission framework, the transmitter necessitates higher energy consumption to support both analog and digital symbols for wireless transmission as compared with the transmitter with single-pattern transmission framework. Also, the widely deployed devices, such as sensor and monitor, have high demand on improving energy efficiency in order to provide a long service. Thus, the efficient energy management and conservation schemes are needed in the DA-TOSC. We should pay more attention to how to improve the energy efficiency of DA-TOSC. Furthermore, how to find an optimal resource allocation scheme among digital coding and analog coding to optimize semantic fidelity is a still open problem in PHY layer. In order to ensure the correct execution of downstream task, how to develop robust schemes to promptly adapt to changes in the PHY policy is very important and still needs to be solved in DA-TOSC. Thus, how to design the HARQ scheme for DA-TOSC framework is an open and interesting research direction. For applications, DA-TOSC is expected to support a wide range of emerging wireless applications. However, the research on DA-TOSC framework is still in its infancy. How to design an effective DA-TOSC framework for supporting the wide range of wireless applications in future wireless networks is very attractive. In future heterogeneous network, different types of device will be equipped with the semantic encoder, which will incorporate the NNs model of different sizes depending on the ability of the device. Due to differences in semantic extraction caused by device diversity, the universal semantic decoding scheme is a challenging research for the DA-TOSC. Also, the robust Hybrid Automatic Repeat reQuest (HARQ) [15], which widely adopted digital systems prevalent in existing networks, can be used for the DA-TOSC framework to reduce the performance effects caused by device diversity.

## VI. Conclusion

In this article, we discussed several challenging problems in deploying TOSC frameworks to the standardized wireless networks. Then, we proposed a digital-analog transmission framework for TOSC to solve these challenging problems. We concluded that our proposed DA-TOSC schemes are expected to efficiently solve the challenging problems in deploying advanced TOSC framework to standardized wireless network, thus opening a wide area for future research. In the future, increased attention should be directed towards several research directions in DA-TOSC.

## References

[1] Y. Yang, J. Wu, T. Chen, C. Peng, J. Wang, J. Deng, X. Tao, G. Liu, W. Li, L. Yang, Y. He, T. Yang, A. H. Aghvami, F. Eliassen, S. Dustdar, D. Niyato, W. Sun, Y. Xu, Y. Yuan, J. Xie, R. Li, and C. Dai, "Task-oriented 6G native-AI network architecture," *IEEE Network*, pp. 1–1, 2023.

[2] W. Weaver, "Recent contributions to the mathematical theory of communication," *The Mathematical Theory of Communication*, 1949.

[3] W. Yang, H. Du, Z. Q. Liew, W. Y. B. Lim, Z. Xiong, D. Niyato, X. Chi, X. Shen, and C. Miao, "Semantic communications for future internet: Fundamentals, applications, and challenges," *IEEE Communications Surveys & Tutorials*, vol. 25, no. 1, pp. 213–250, 2023.

[4] ITU-T, "Requirements and reference architecture of semantic-aware networks in future networks," in *Recommendation*, 2023. [Online]. Available: https://www.itu.int/itu-t/workprog/wp_item.aspx?isn=19137

[5] X. Kang, B. Song, J. Guo, Z. Qin, and F. R. Yu, "Task-oriented image transmission for scene classification in unmanned aerial systems," *IEEE Transactions on Communications*, vol. 70, no. 8, pp. 5181–5192, 2022.

[6] D. Huang, F. Gao, X. Tao, Q. Du, and J. Lu, "Toward semantic communications: Deep learning-based image semantic coding," *IEEE Journal on Selected Areas in Communications*, vol. 41, no. 1, pp. 55–71, 2023.

[7] Y. Fu, W. Cheng, W. Zhang, and J. Wang, "Scalable extraction based semantic communication for 6G wireless networks," *IEEE Communications Magazine*, pp. 1–7, 2023.

[8] S. Ma, W. Qiao, Y. Wu, H. Li, G. Shi, D. Gao, Y. Shi, S. Li, and N. Al-Dhahir, "Task-oriented explainable semantic communications," *IEEE Transactions on Wireless Communications*, vol. 22, no. 12, pp. 9248–9262, 2023.

[9] M. Jankowski, D. Gündüz, and K. Mikolajczyk, "Wireless image retrieval at the edge," *IEEE Journal on Selected Areas in Communications*, vol. 39, no. 1, pp. 89–100, 2021.

[10] H. Xie and Z. Qin, "A lite distributed semantic communication system for internet of things," *IEEE Journal on Selected Areas in Communications*, vol. 39, no. 1, pp. 142–153, 2021.

[11] X. Jiang and H. Lu, "Joint rate and resource allocation in hybrid digital-analog transmission over fading channels," *IEEE Transactions on Vehicular Technology*, vol. 67, no. 10, pp. 9528–9541, 2018.

[12] C. Lan, C. Luo, W. Zeng, and F. Wu, "A practical hybrid digital-analog scheme for wireless video transmission," *IEEE Transactions on Circuits and Systems for Video Technology*, vol. 28, no. 7, pp. 1634–1647, 2018.

[13] H. Xie and Z. Qin, "A lite distributed semantic communication system for internet of things," *IEEE Journal on Selected Areas in Communications*, vol. 39, no. 1, pp. 142–153, 2021.

[14] B. Tan, J. Wu, R. Wang, W. Luo, and J. Liu, "An optimal resource allocation for hybrid digital-analog with combined multiplexing," *IEEE Internet of Things Journal*, vol. 6, no. 1, pp. 1125–1135, 2019.

[15] P. Jiang, C.-K. Wen, S. Jin, and G. Y. Li, "Deep source-channel coding for sentence semantic transmission with HARQ," *IEEE Transactions on Communications*, vol. 70, no. 8, pp. 5225–5240, 2022.

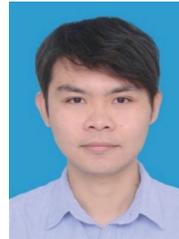

**Yuzhou Fu** received the M.S. degree in school of computer, electronics and information from Guangxi University, Guangxi, China, in 2019. He is currently pursuing a Ph.D. degree in telecommunication engineering at Xidian University. His research interests include semantic communication and maritime wireless communication network optimization.

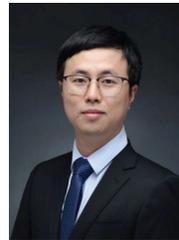

**Wenchi Cheng** received his B.S. and Ph.D. degrees in telecommunication engineering from Xidian University in 2008 and 2013, respectively, where he is a full professor. He was a visiting scholar with the Department of Electrical and Computer Engineering, Texas A&M University, College Station, from 2010 to 2011. His current research interests include B5G/6G wireless networks, emergency wireless communications, and OAM based wireless communications. He has published more than 100 international journal and conference papers in IEEE JSAC, IEEE magazines, and IEEE transactions, and at conferences including IEEE INFOCOM, GLOBECOM, ICC, and more. He has served or is serving as an Associate Editor for the IEEE Systems Journal, IEEE Communications Letters, and IEEE Wireless Communications Letters, as the Wireless Communications Symposium Co-Chair for IEEE ICC 2022 and IEEE GLOBECOM 2020, the Publicity Chair for IEEE ICC 2019, and the Next Generation Networks Symposium Chair for IEEE ICCC 2019.

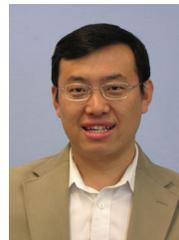

**Wei Zhang** received the Ph.D. degree in electronic engineering from The Chinese University of Hong Kong in 2005. He is currently a Professor with the School of Electrical Engineering and Telecommunications, University of New South Wales, Sydney, NSW, Australia. He has published more than 200 articles and holds five U.S. patents. His research interests include millimetre wave communications and massive MIMO. He is the Vice Director of the IEEE ComSoc Asia Pacific Board. He serves as an Area Editor for the IEEE TRANSACTIONS ON WIRELESS COMMUNICATIONS and the Editor-in-Chief for *Journal of Communications and Information Networks*.

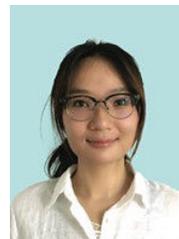

**Jingqing Wang** received the B.S. degree from Northwestern Polytechnical University, Xi'an, China, in Electronics and Information Engineering and the Ph.D. degree in Electrical and Computer Engineering from Texas A&M University, College Station, TX, USA. She is currently an Assistant Professor at Xidian University. She won the Best Paper Award from the IEEE GLOBECOM in 2020 and 2014, respectively, the Hagler Institute for Advanced Study Heep Graduate Fellowship Award from Texas A&M University in 2018, and Dr. R.K. Pandey and Christa U. Pandey'84 Fellowship, Texas A&M University, USA, 2020-2021. Her research interests focus on B5G/6G mobile wireless network technologies, statistical QoS provisioning, 6G mURLLC, information-theoretic analyses of FBC, and emerging machine learning techniques over 5G and beyond mobile wireless networks.